\documentclass[conference, a4paper]{IEEEtran}
\usepackage[left=1.59cm,right=1.59cm,top=1.91cm,bottom=2.54cm]{geometry}
\IEEEoverridecommandlockouts
\usepackage{cite}
\usepackage{amsmath,amssymb,amsfonts}
\usepackage{algorithmic}
\usepackage{graphicx}
\usepackage{textcomp}
\usepackage{xcolor}
\usepackage{soul}

\usepackage{eurosym}
\usepackage{amstext} 
\DeclareRobustCommand{\officialeuro}{%
  \ifmmode\expandafter\text\fi
  {\fontencoding{U}\fontfamily{eurosym}\selectfont e}}

\def\BibTeX{{\rm B\kern-.05em{\sc i\kern-.025em b}\kern-.08em
    T\kern-.1667em\lower.7ex\hbox{E}\kern-.125emX}}
\begin{document}

\title{
Management of Electric Vehicles as Flexibility Resource for Optimized Integration of Renewable Energy with Large Buildings \\
\thanks{Support for this research was provided by the Fundação para a Ciência e a Tecnologia (Portuguese Foundation for Science and Technology) through the Carnegie Mellon Portugal Program.}
}

\author{\IEEEauthorblockN{Pedro Moura}
\IEEEauthorblockA{\textit{Electrical and Computer Eng.} \\
\textit{University of Coimbra}\\
Coimbra, Portugal \\
pmoura@isr.uc.pt}

\and
\IEEEauthorblockN{Greta K.W. Yu}
\IEEEauthorblockA{\textit{Carnegie Institute of Tech.} \\
\textit{Carnegie Mellon University}\\
Pittsburgh, USA \\
kaiweiy@andrew.cmu.edu}
\and
\IEEEauthorblockN{Javad Mohammadi}
\IEEEauthorblockA{\textit{Electrical and Computer Eng.} \\
\textit{Carnegie Mellon University}\\
Pittsburgh, USA \\
jmohamma@andrew.cmu.edu}

}

\maketitle

\begin{abstract}
The increasing penetration of renewable generation requires new tools to achieve a high matching between demand and renewable generation at building and community levels. Large commercial and public buildings with parking lots have a considerable potential to provide energy flexibility by controlling the charging of Electric Vehicles (EVs) and injecting part of the stored energy into the building, using Building-to-Vehicle (B2V) and Vehicle-to-Building (V2B) systems. However, EVs and buildings do not often belong to the same entity and in Portugal the existing regulation does not allow financial transactions between buildings and EVs as separate entities. Addressing this regulation hurdle requires innovative optimization methods for the implementation of B2V/V2B systems. Moreover, the new legislation regarding the self-consumption of renewable generation in Portugal enables the trade of renewable generation surplus between buildings and the establishment of renewable energy communities. This paper intends to address this issue and proposes a formulation to aggregate and manage the sharing of generation surplus between buildings, using EVs as a flexibility resource. The simulation results showcase the achieved increase in renewable self-consumption at building and community levels, as well as the reduction in electricity costs.
\end{abstract}


\section*{Nomenclature}
\subsection*{Inputs}
\addcontentsline{toc}{section}{Nomenclature}
\begin{IEEEdescription}[\IEEEusemathlabelsep\IEEEsetlabelwidth{$V_1,V_2,V_3$}]
\item[$C_P$]  Baseline parking tariff for EVs ($\euro{}/h$)
\item[$C_C(h),$] Tariff for the charging/discharging of EVs at 
\item[$C_D(h)$] \;\ time step $h$ ($\euro{}/h$)
\item[$C_F$] Reward for EV charging flexibility ($\euro{}/h$)
\item[$C_{EG}(h),$] Tariff for power exported/imported to/from 
\item[$C_{IG}(h)$] \;\ the grid at time step $h$ ($\euro{}/kWh$)
\item[$C_{G}(h)$] Tariff for the grid use between buildings
\item[$ $] \;\ in the community at time step $h$ ($\euro{}/kWh$)
\item[$t^{b+}_{R,n},$] Charging/discharging period in building $b$ 
\item[$t^{b-}_{R,n}$] \;\ requested by EV owner $n$ ($hour$)
\item[$t^b_{P,n}$] Total parking period of EV $n$ in building $b$ ($h$)
\item[$L^b(h)$] Net electricity load in building $b$, excluding  \item[$ $] \;\  the impact of EVs, at time step $h$ ($kW$)
\item[$P_{EV,n}^{b+<}$,] Maximum charging/discharging power for EV 
\item[$P_{EV,n}^{b-<}$] \;\ $n$ in building $b$ ($kW$)
\item[$\eta^b_n$] Efficiency of the EV charging/discharging for 
\item[$ $] \;\ for EV $n$  parked in building $b$ ($\%$)
\end{IEEEdescription}

\subsection*{Variables}
\addcontentsline{toc}{section}{Nomenclature}
\begin{IEEEdescription}[\IEEEusemathlabelsep\IEEEsetlabelwidth{$V_1,V_2,V_3$}]
\item[$C_E^b(h)$] Electricity cost of building $b$ at time step 
\item[$ $] \;\ $h$ ($\euro{}$)
\item[$C^b_{EV}(n)$] Total cost of EV $n$ parked in building $b$ ($\euro{}$)
\item[$C_{EC}(h),$] Tariff for power exported/imported to/from 
\item[$C_{IC}(h)$] \;\ the community at time step $h$ ($\euro{}/kWh$)
\item[$t^{b+}_{T,n},$] Total charging/discharging period of EV $n$ in 
\item[$t^{b-}_{T,n}$] \;\ building $b$ ($h$)
\item[$t^{b+}_{U,n}(h)$,] Net used charging/discharging period of EV 
\item[$t^{b-}_{U,n}(h)$] \;\ $n$ at time step $h$ in building $b$ ($h$)
\item[$P_{EV,n}^{b+}(h),$] Charging/discharging power of EV $n$ parked 
\item[$P_{EV,n}^{b-}(h)$] \;\ in building $b$ at time step $h$ ($kW$)
\item[$P^{b+}_c(h)$,] Exporting/importing power flow at time step 
\item[$P^{b-}_c(h)$] \;\ $h$  between building $b$ and community $c$ ($kW$)
\item[$P^{b+/-}_g$] Power flow between the building $b$ and the
\item[$ $] \;\ grid $g$ ($kW$)
\item[$P^{b+/-}_c$] Power flow between the building $b$ and the
\item[$ $] \;\ community $c$ ($kW$)
\end{IEEEdescription}

\section{Introduction}
\subsection{Motivation}
With 55\% of the electricity generation in 2018 ensured by renewable resources, Portugal already has a large share of renewables, leading to periods with renewable generation surplus due to the intermittent nature of the resources \cite{Delgado2018}. Additionally, Portugal has ambitious targets aiming at achieving 100\% renewable electricity generation by 2050, with 50\% of the total installed capacity ensured by solar photovoltaic (PV) generation, being 50\% of such capacity ensured by decentralized generation, mainly installed in buildings \cite{RNC2050}. However, in most buildings, there is a high mismatch between the local PV generation and demand profiles, leading to the injection of large share of the generation into the grid, causing technical challenges and financial losses. 

The Portuguese government also aims at achieving 70\% electrification of transports by 2050 \cite{RNC2050}. Electric Vehicles (EVs) have the potential to provide the needed flexibility requirements for the integration of distributed generation in buildings, adjusting the charging period using the Building-to-Vehicle (B2V) system, or injecting into the building part of the stored energy using the Vehicle-to-Building (V2B) system \cite{moura2020linking}.
Hence, if properly managed, large buildings with parking lots offer a large flexibility capacity. However, typically, EVs and buildings do not belong to the same entity and the Portuguese regulation does not allow electricity trading between buildings and EV users. Therefore, this paper intends to unlock the flexibility potential of B2V and V2B. 

On the other hand, the new legislation for the self-consumption of renewable generation in Portugal enables bilateral contracts between buildings or the establishment of renewable energy communities, in order to trade the renewable generation surplus. Therefore, an aggregated optimization at the community level will be needed for the establishment of a renewable energy community. A central aggregator must be considered to coordinate the share of generation surplus between buildings, using EVs as a flexibility resource, in order to ensure the matching between generation and consumption.

\subsection{Related Works}
There is a vast body of works proposing methodologies to implement V2B strategies, the aggregation of buildings and their economic relationship.

The management of EVs is mainly considered in residential buildings. Authors in \cite{Gray2018} present a transactive energy control for residential prosumers with batteries and EVs and
in \cite{Liu2019b} a two-stage optimal charging scheme based on transactive control is proposed to manage day-ahead electricity procurement and real-time EV charging management. 
However, in residential buildings, the EV and the building belong to the same entity and there are no economic transactions between buildings and EVs.

There are also several works focused on the integration of EVs in office buildings. For example, \cite{Jin2017} considers an office building with flexible demand and EVs with the objective of ensuring load leveling and \cite{Thomas2018} focuses an office building equipped with PV, storage and EVs that aims to minimize the total energy costs. However, these papers do not consider the need for establishing an economic relationship between building and EVs although they belong to different entities.

The economic relationship between EV users and buildings is explored by some researchers. Reference \cite{Nefedov2018} considers an office building with PV and EVs with the objective of minimizing energy costs and 
\cite{Quddus2018} considers several commercial buildings and EV charging stations with the objective of minimizing the costs of energy in the building and the charging costs. In \cite{Liu2019} a transactive real-time EV charging management scheme is proposed for the building energy management system of commercial buildings with PV on-site generation and EV charging services. However, such works assume that buildings and EVs can trade electricity which does not comply with existing regulation in most countries. Additionally, the proposed formulations usually require complex information from the EV users that is not easily obtained in real scenarios.

\subsection{Contribution}
\begin{figure}[htbp]
\centerline{\includegraphics[trim=0 0 0 0,clip, width=0.45\textwidth] {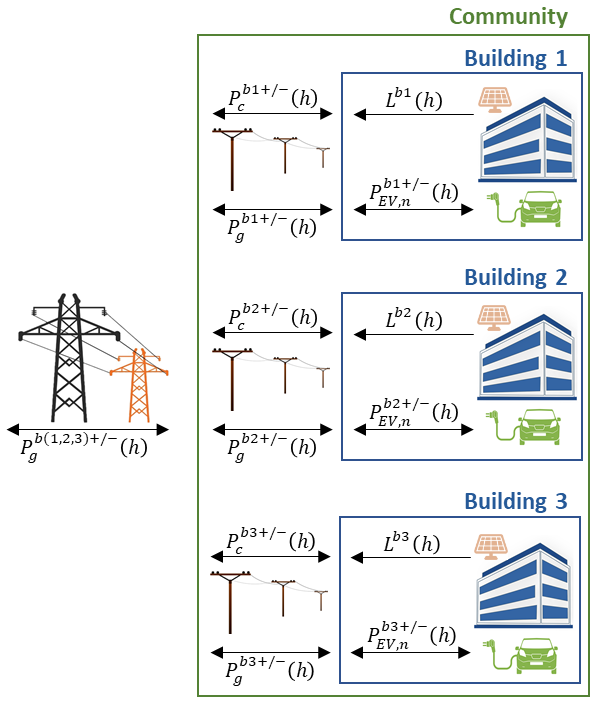}}
\caption{Power flows in the renewable energy community}
\label{Community}
\end{figure}

This paper introduces a renewable energy community, constituted by large commercial and public buildings, with a central aggregator aligned with the new legislation for the self-consumption of renewable generation in Portugal. As presented in Fig.~\ref{Community}, each building can sell the generation surplus to the grid or to other buildings in the community, as well as buy electricity from the grid or from the community. The buildings do not have to be located in the same neighborhood, since the electrical grid can be used, being paid the tariff for grid access for the used voltage levels. Each building can only sell the generation surplus to the community if another building intends to buy it, otherwise, such surplus must be injected into the grid.

\vspace{-0.2cm}

The formulation implements a market between the buildings of the community determining the tariffs for the transactions of renewable generation surplus in the community, being such tariffs in direct competition with the grid tariffs. The aggregation ensures the sharing of generation surplus between buildings, using EVs as a flexibility resource, in order to ensure the matching between generation and demand in the community. Additionally, the large-scale integration of renewables into the grid is promoted by a tariff proportional to the variation of the wholesale market that is strongly influenced by the renewable generation levels. Therefore, the paper uses economic techniques to manage the power flow between buildings in the community and between the community and the grid, establishing a transactive energy control that can be implemented with the actual legislation in Portugal. 

Moreover, in the EVs flexibility management, the formulation connects the value of electricity and parking duration to design a novel V2B methodology that can be implemented with the existing legislation in Portugal. The EV users only have to provide information about the intended time of departure (allowing to define the associated parking period), as well as the desired charging and maximum discharging periods. Such an approach simplifies the implementation, since with the use of a charging period, instead of the State of Charge (SoC), the monitoring of EV data is not required and the use of a discharging period allows the users to control (or avoid) the discharging of their batteries. If the user does not select any charging and discharging period, before each parking use, such EV is not charged. Instead of electricity transactions, the methodology is based on the parking time, being the EV charging considered as an added-value service, which cost can be reduced if the user allows charging flexibility for B2V and discharging for V2B. Therefore, with such strategy, there is no electricity trading between the building and EVs, and there is no injection of electricity from the discharging of EVs into the grid, therefore complying with the actual legislation.

\subsection{Paper Organization}
The remainder of the paper is structured as follows. Section~II presents the problem formulation. Section~III presents the data used for the test cases and Section~IV presents the simulation results. Finally, Section~V summarizes the paper, emphasizing its main conclusions.

\section{Problem Formulation}

\subsection{Objective Function}
The proposed problem aims at minimizing the total costs from the community perspective, considering $B$ buildings, during all time steps (i.e., $H$). The objective function \eqref{eq:obj} accounts for the electricity costs in each building, as well as the profit associated with the parking, charging and discharging of $N$ EVs parked at each building $b$.

\color{black}
\small
\begin{equation}
\textrm{min}  \sum_{b=1}^{B} \left(\sum_{h=1}^{H}C^b_E(h)-\sum_{n=1}^{N}C^b_{EV}(n)\right)
\label{eq:obj}
\end{equation}
\normalsize

The net cost of the electricity consumption and self-generation \eqref{eq:ce} in each building $b$ consists of four parts: (i) the cost of energy drawn from the community; (ii) the financial compensation for the energy injected into the community; (iii) the cost of energy drawn from the grid; and (iv) the financial compensation for the energy injected into the grid. Note, $\mathbb{P}[.]_{+}$ and $\mathbb{P}[.]_{-}$ are operators that preserve only positive and negative values, respectively.

\vspace{-0.3cm}
\small
\begin{multline}
C_E^b(h)=\Delta h \left(
P^{b-}_c(h) \cdot C_{IC} - P^{b+}_c(h) \cdot C_{EC} \right.+
\\
\mathbb{P}\left[ L^b(h)-P^{b-}_c(h)+\sum_{n=1}^{N}P_{EV,n}^{b+}(h)-\sum_{n=1}^{N}P_{EV,n}^{b-}(h) \right]_{-} \cdot C_{IG}+
\\
\left.
\mathbb{P}\left[ L^b(h)-P^{b+}_c(h)+\sum_{n=1}^{N}P_{EV,n}^{b+}(h)-\sum_{n=1}^{N}P_{EV,n}^{b-}(h) \right]_{+} \cdot C_{EG}\right)
\label{eq:ce}
\end{multline}
\normalsize

The time references that are submitted by the EV owner $n$ before entering the parking lot are the: requested parking period; requested charging period; and allowed discharging period. However, the used discharging period can be lower than the value allowed by the user and the total charging period can be higher than the requested value in order to compensate any used discharging period. Therefore, the total parking costs \eqref{eq:ct}, for each EV $n$ in building $b$, depend on the parking period, used periods for charging and discharging in each time step, and the total charging and discharging periods over all time steps. Equations \eqref{eq:tcu} and \eqref{eq:tdu} derive the net used charging and discharging periods in each time step and calculate the total periods over all time steps (i.e., $H$). The discharging tariff should be defined to ensure a profit higher than the cost associated with the battery degradation.

\vspace{-0.3cm}
\small
\begin{multline}
C^b_{EV}(n)=t^b_{P,n} \cdot C^b_P  +  (t^b_{P,n} - t^{b+}_{T,n} - t^{b-}_{T,n} ) \cdot C_F \\
+ \sum_{h=1}^{H} \left( t^{b+}_{U,n}(h)  \cdot C_C(h) \right) +
\sum_{h=1}^{H} \left(t^{b-}_{U,n}(h) \cdot C_D(h)  \right) 
\label{eq:ct}
\end{multline}
\vspace{-0.7cm}

\begin{equation}
t^+_{U,n}(h)= \mathbb{P} \left[P_{EV,n}^+(h)\right]_+ \cdot \frac{\Delta h}{P_{EV,n}^{+<}}, \;\ t^+_{T,n}= \sum_{h=1}^{H} t^+_{U,n}(h)
\label{eq:tcu}
\end{equation}

\vspace{-0.2cm}

\begin{equation}
t^-_{U,n}(h)= \mathbb{P} \left[P_{EV,n}^-(h)\right]_- \cdot \frac{\Delta h}{P_{EV,n}^{-<}}, \;\ t^+_{T,n}= \sum_{h=1}^{H} t^+_{U,n}(h)
\label{eq:tdu}
\end{equation}
\normalsize

\subsection{Constraints}
The defined objective is subject to constraints related with the required  periods, as well as with the charging and discharging power, power flows and costs. The charging period achieved until the end of the parking period \eqref{eq:tct} should be enough to ensure the satisfaction of the charging period requested by the user and to compensate for the used discharging period, including the losses. Additionally, since the requested periods were defined based on the maximum power, the charging and discharging periods must be corrected by the ratio between the average and the maximum power.

\small
\begin{equation}
t^{b+}_{T,n}=t^{b+}_{R,n}\frac{P_{EV,n}^{b+<}}{\overline{P}_{EV,n}^{b+}}+\frac{t^{b-}_{T,n}}{\eta^b_n}\frac{ \overline{P}_{EV,n}^{b-}}{P_{EV,n}^{b-<}} \label{eq:tct}
\end{equation}
\normalsize

The total discharging period must be lower than the maximum period allowed by the user and the used discharging period until the actual time step $x$ must be lower than the used charging period \eqref{eq:tdm}, in order to ensure that a SoC lower than the initial value is never achieved (eliminating the risk of reaching the minimum SoC). The required parking and charging periods defined by the user must be positive and the allowed discharging period cannot be negative \eqref{eq:tp}. The charging and discharging is also limited by the charging infrastructure to the maximum power \eqref{eq:ch-limit}. 

\small

\begin{equation}
t^{b-}_{T,n} \leq t^{b-}_{R,n}, \;\ \sum_{h=1}^{x} t^{b-}_{U,n}(h) < \sum_{h=1}^{x} t^{b+}_{U,n}(h) \label{eq:tdm}
\end{equation}


\begin{equation}
\mathfrak{t}_{P,n}>0, \;\ t^+_{R,n}>0, \;\ t^-_{R,n}\geq 0 \label{eq:tp}
\end{equation}


\begin{equation}
0\leq P_{EV,n}^{b+}(h) \leq P_{EV,n}^{b++}, \;\ 0\leq P_{EV,n}^{b-}(h) \leq P_{EV,n}^{b--}
\label{eq:ch-limit}\\
\end{equation}
\normalsize

The power flow between each building and the community can only have one direction in each time slot \eqref{eq:pb} and the import or export power flow between the building and the community is limited to the net load of such building added by the impact of charging and discharging the parked EVs \eqref{eq:pcmax}. It is only possible to export to the community if other building needs to import such energy, therefore the power flows in the community must cancel each other \eqref{eq:pc}.

\small
\begin{equation}
P^{b+}_c(h) \cdot P^{b-}_c(h) = 0
\label{eq:pb}
\end{equation}

\vspace{-0.3cm}

\begin{multline}
|P^{b+}_c(h) + P^{b-}_c(h)| \leq
\\|L^b(h)+\hspace{-0.1cm}\sum_{n=1}^{N}P_{EV,n}^{b+}(h)-\hspace{-0.1cm}\sum_{n=1}^{N}P_{EV,n}^{b-}(h)|
\label{eq:pcmax}
\end{multline}


\begin{equation}
\sum_{b=1}^{B} \left(P^{b+}_c(h) + P^{b-}_c(h) \right) = 0
\label{eq:pc}
\end{equation}
\normalsize

In order to provide incentives to share the renewable generation surplus in the community, the tariff for exporting energy to the community must be lower than the tariff of exporting it to the grid \eqref{eq:cec}, since the tariff for the exported energy have negative values. Simultaneously, the cost of importing energy from the community should not be higher than the equivalent cost from the grid and ensure the compensation of the grid utilization between the two buildings \eqref{eq:cic}.

\small
\begin{equation}
C_{EC}(h) \leq  C_{EG}(h) 
\label{eq:cec}
\end{equation}

\vspace{-0.5cm}

\begin{equation}
C_{IC}(h) \leq C_{IG}(h), \;\ C_{IC}(h) = |C_{EC}(h)|+ C_{G}(h)
\label{eq:cic}
\end{equation}
\normalsize

\vspace{-.1cm}

\section{Data and Scenarios}
\subsection{Buildings}
The simulations use data from the Department of Electrical and Computer Engineering at the University of Coimbra (Portugal). The building has a total area of about 10,000~$m^2$, electricity consumption of about 500~$MWh/year$ and a PV system with 79~$kWp$, which covers about 16\% of the existing electricity demand \cite{Fonseca2018}. The PV generation was adjusted for a future scenario ensuring 50\% of the demand, in order to have periods with renewable generation surplus. The used data is from March, a month of intermediate consumption and generation. In order to provide data for different buildings, but ensuring the same type of building, data from four days of different weeks in March from the same building were selected to represent four different buildings. 

\begin{figure}[htbp]
\centerline{\includegraphics[trim=0 3 10 16,clip, width=0.5\textwidth] {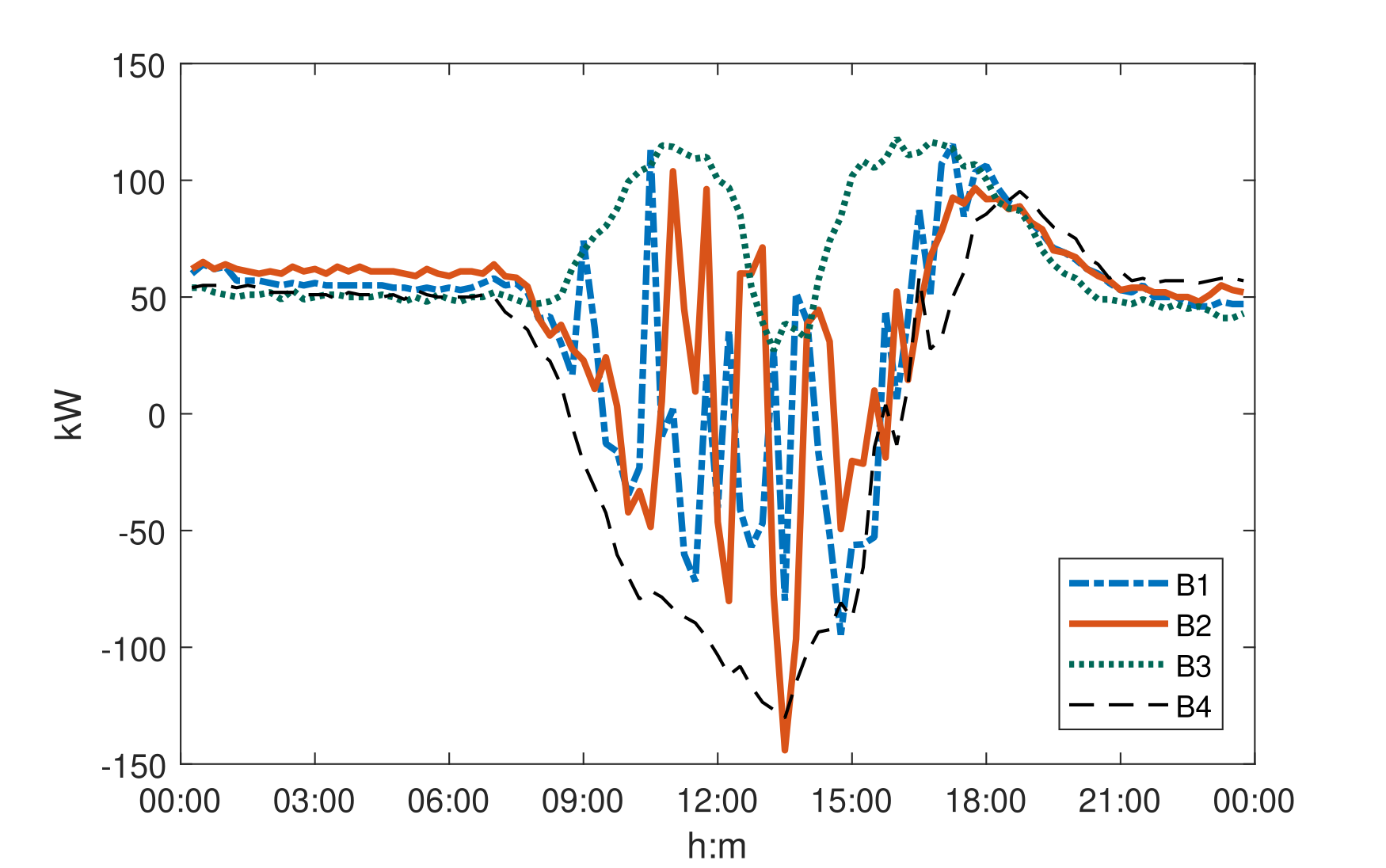}}
\caption{Net load in the four considered buildings}
\label{netload}
\end{figure}

Fig.~\ref{netload} presents the net loads considered for the four buildings. Being the input the net load, and not directly the PV generation, the generation surplus does not have a proportional variation in the different buildings. Therefore, it can be assumed that, in an average scenario, there are simultaneously buildings with surplus and others with a deficit of PV generation. It was therefore considered a community of buildings in one University campus. 

It was considered a tariff for the electricity consumed from the grid with an average cost equal to the actual average cost of the tariff in the reference building (122.8~$\euro{}/MWh$), but with a variation proportional to the wholesale market, being selected data from a typical day in March. For the tariff for the electricity exported to the grid, it was considered a flat tariff with 90\% of the monthly average of the wholesale market (-35.8~$\euro{}/MWh$), as defined by the actual legislation. It was also considered a flat tariff of 50~$\euro{}/MWh$ for the grid use between buildings. For the EVs, the parking, flexibility and discharging considered flat tariffs of 0.5~$\euro{}/h$, -0.5~$\euro{}/h$ and -3~$\euro{}/h$, respectively, being used for the charging a tariff with an average cost of 2~$\euro{}/h$ and a variation proportional to the tariff for the electricity consumed from the grid.


\subsection{Electric Vehicles}
The simulations considered EVs available in the buildings between 8~$a.m.$ and 8~$p.m.$ with an average of 8~$hours$, 2~$hours$ and 0.75~$hours$ for the parking, charging and discharging periods, respectively. The requirements of 30 EVs were generated with a small standard deviation (1, 0.5 and 0.25 for the parking, charging and discharging, respectively) in order to ensure uniform requirements. Such periods are aligned with the typical use of parking in University campuses. For each building, six EVs were randomly selected from the 30 available EV profiles. The used chargers have a maximum charging/discharging power of 10~$kW$ and 93\% of efficiency.


\section{Simulation Results}
\begin{figure}[htbp]
\centerline{\includegraphics[trim=0 3 10 16,clip, width=0.5\textwidth] {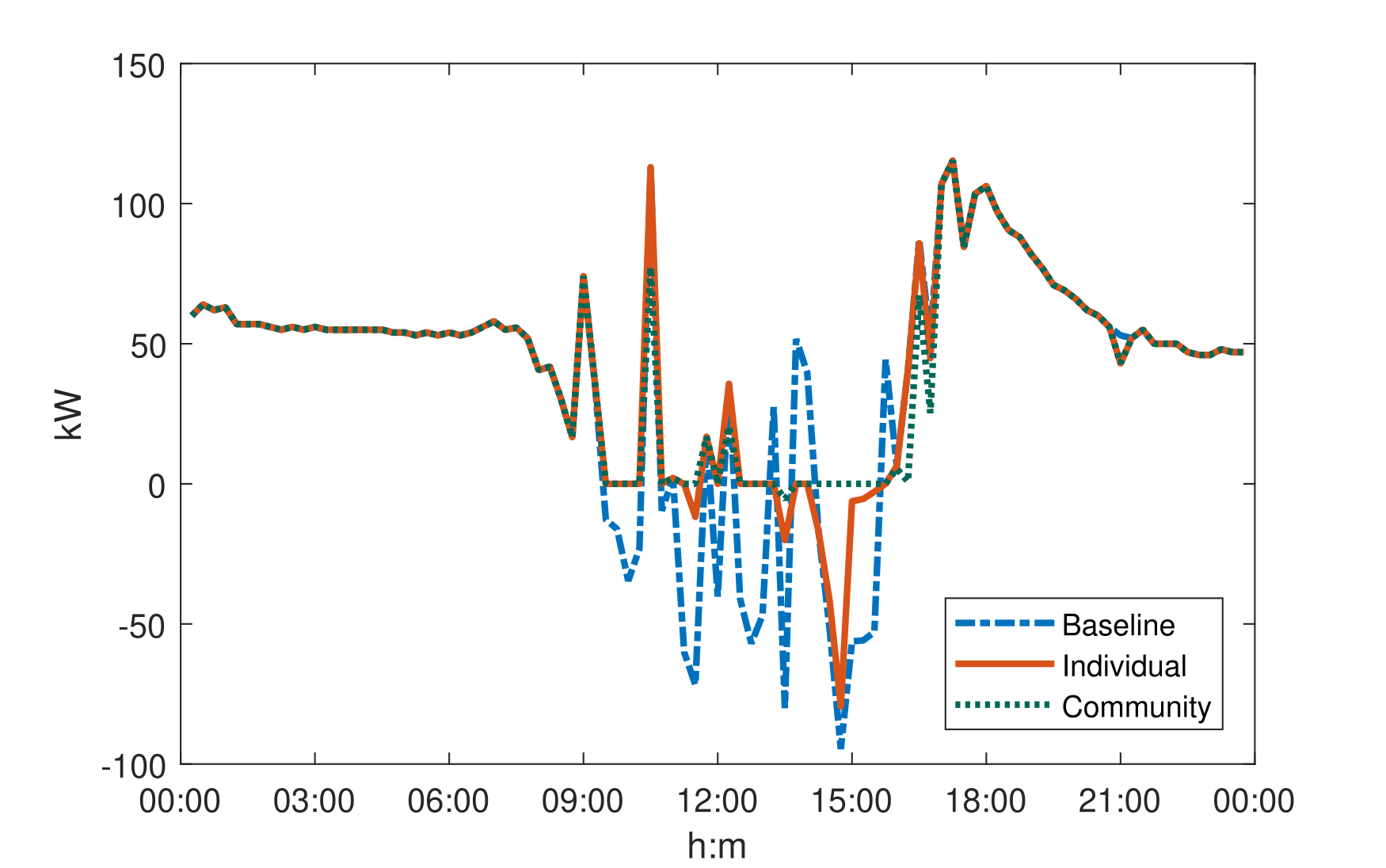}}
\caption{Net load for building 1 with individual and community management}
\label{NetLoad1}
\end{figure}

The formulation was implemented in Python using Gurobi Optimizer as linear optimization solver. Fig.~\ref{NetLoad1} presents the net load for the baseline, and for two scenarios with EVs (with individual community management of buildings). As can be seen, the EVs preferentially charge and discharge in periods of negative (generation surplus) and positive (generation deficit) net load, respectively. The presented net load is from the grid point-of-view, not including the power flow between the building and the community. This justifies the different profiles for the individual and community management, since in the community management the community acts as an additional flexibility resource, minimizing the periods with negative net load and slightly changing the charging/discharging profile.

\begin{figure}[htbp]
\centerline{\includegraphics[trim=0 3 10 16,clip, width=0.5\textwidth] {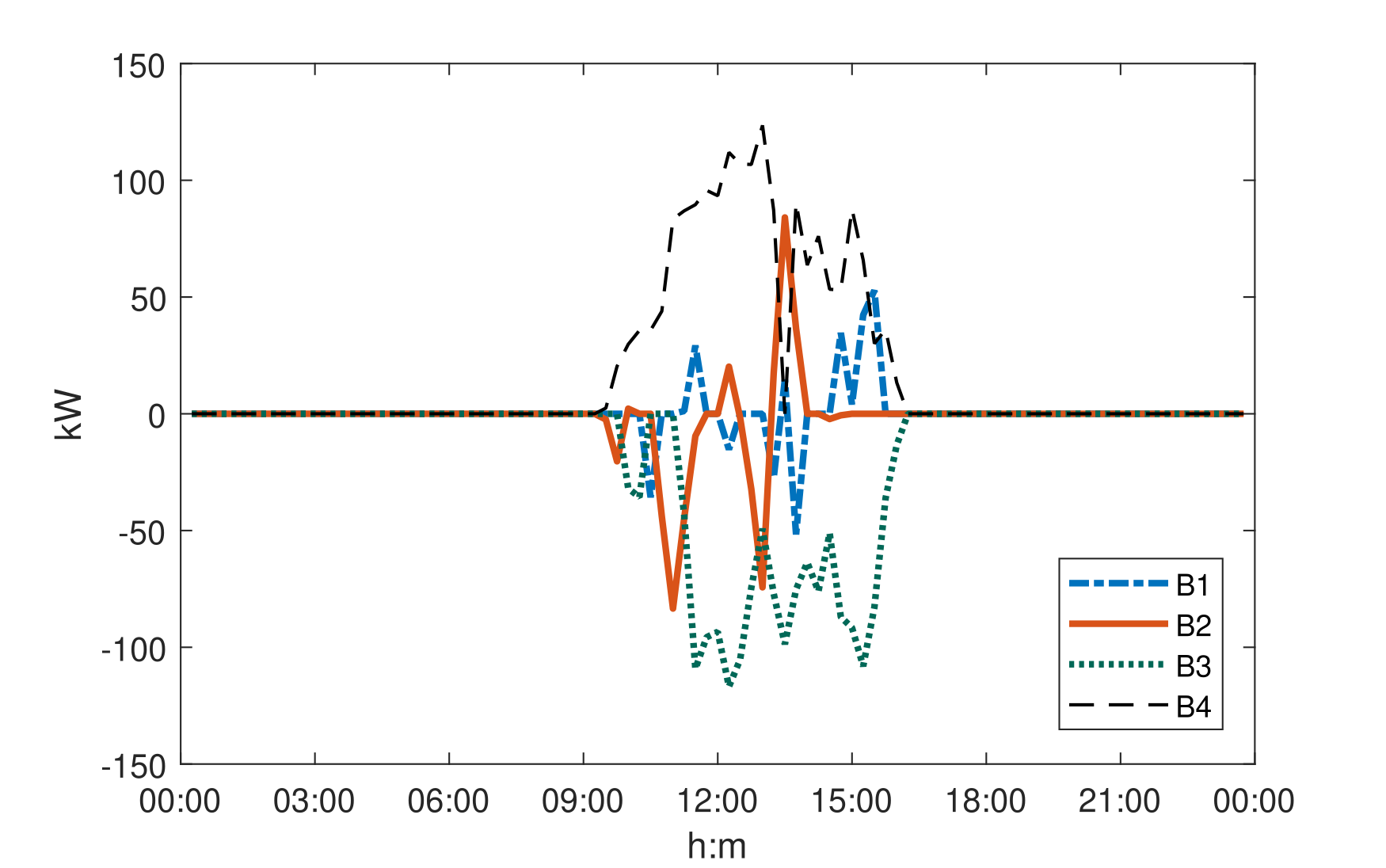}}
\caption{Power flow between each building and the community}
\label{PF}
\end{figure}

The other three buildings were also simulated with individual and community management. Fig.~\ref{PF} presents the power flow between each building and the community. As can be seen, building 4 and building 1 only export and import energy, respectively, while building 2 and 3 simultaneously export and import during different periods of the day. As result of the market in the community, the tariffs for the energy exported to and imported from the community were -35.8~$\euro{}/MWh$ and 85.8~ $\euro{}/MWh$, respectively, therefore, ensuring an exporting tariff equal to the tariff offered by the grid, but with a much lower tariff for the importing tariff.

Tab.~\ref{tab:Costs} presents the costs achieved in the several simulated scenarios, considering the electricity costs in the building, the costs of parking and charging EVs in such building paid by the users, as well as the total costs in the building, including electricity and EVs costs (objective function). Including the EVs in each building (individual scenario) the electricity costs increase due to the higher demand required for the  charging. However, due to the use of charging and discharging flexibility, it was possible to achieve an electricity consumption increase of 12.3\% with a cost increase of only 4.6\%. Additionally, by considering community management it was possible to reduce the electricity costs relative to independent management by 3.9\%. When considering the costs paid by EV users, the total costs relative to the baseline were reduced in 19\% and 22.3\% for independent and community management, respectively. 

\vspace{-0.2cm}
\small
\begin{table}[h]
\centering
\renewcommand{\arraystretch}{1.3}
\caption{Costs by building and scenario ($\euro{}$)}
\label{tab:Costs}
\centering
\begin{tabular}{c | r | r r r | r r r}
Buil. & \multicolumn{1}{c|}{Base.} & \multicolumn{3}{c|}{Individual} & \multicolumn{3}{c}{Community} \\
 \# & \multicolumn{1}{c|}{$C_{E}$} & \multicolumn{1}{c}{$C_{E}$} & \multicolumn{1}{c}{$C_{EV}$} & \multicolumn{1}{c|}{Obj.} & \multicolumn{1}{c}{$C_{E}$} & \multicolumn{1}{c}{$C_{EV}$} & \multicolumn{1}{c}{Obj.} \\
\hline
1 & 129.6 & 129.8 & -30.3 & 99.5 & 128.4 & -30.0 & 98.4\\
2 & 140.0 & 147.1 & -38.2 & 108.9 & 143.3 & -38.0 & 105.2\\
3 & 201.9 & 218.5 & -33.0 & 185.5 & 200.9 & -31.3 & 169.6\\
4 & 90.0 & 92.1 & -31.0 & 61.2 & 92.1 & -29.0 & 63.2\\
\hline
Total & 561.5 & 587.5 & -132.4 & 455.0 & 564.6 & -128.3 & 436.4\\
\end{tabular}
\end{table}
\normalsize
\vspace{-.2cm}

\section{Conclusion}
This paper proposes a method to aggregate flexibility potential of large commercial and public buildings in a renewable energy community setup. The proposed method is aligned with the new Portuguese self-consumption legislation, to ensure the sharing of generation surplus between buildings with the objective of achieving the high matching between demand and renewable generation at a community level. Such community uses EVs as a flexibility resource and proposes a mathematical formulation that enables the implementation of V2B/B2V systems while following the current Portuguese legislation that does not allow electricity transactions between EV users and buildings. The formulation was simulated considering four buildings, showing that the proposed solution results in maximizing matching between local renewable generation and electricity demand, using the flexibility provided by the charging and discharging of EVs, as well as the import and export to the community. The results also show a higher reduction of costs and a higher self-consumption of local renewable generation achieved with the community management when compared with the individual management of each building, therefore, highlighting the effectiveness of the proposed formulation. We intend to extend this work by using our recent research work on collaborative learning \cite{mohammadi2019collaborative} and test the outcomes using our campus wider CarnegiePLUG test-bed. 

\bibliographystyle{./bibliography/IEEEtran}
\bibliography{./bibliography/IEEEabrv,./bibliography/IEEEexample}

\end{document}